\documentclass[preprint2]{aastex}

\usepackage{graphicx}

\begin{document}


\title{X-ray high-resolution spectroscopy reveals feedback in a Seyfert galaxy from an ultra fast wind with complex ionization and velocity structure}

\author{A.L. Longinotti\altaffilmark{1,2,3}, Y. Krongold\altaffilmark{2},  M. Guainazzi\altaffilmark{3,4}, M. Giroletti\altaffilmark{5}, F. Panessa\altaffilmark{6},  E.~Costantini\altaffilmark{7}, M. Santos-Lleo\altaffilmark{3}, P.~Rodriguez-Pascual\altaffilmark{3}}
\affil{1 Catedr\'atica CONACYT - Instituto Nacional de Astrof\'isica, \'Optica y Electr\'onica, Luis E. Erro 1, Tonantzintla, Puebla, M\'exico, C.P. 72840}
\affil{2 Instituto de Astronomia, Universidad Nacional Autonoma de Mexico, Apartado Postal 70264, 04510 Mexico D.F., Mexico}
\affil{3 ESAC, P.O. Box, 78 E-28691 Villanueva de la Ca\~nada, Madrid, Spain}
\affil{4 Institute of Space and Astronautical Science, 3-1-1 Yoshinodai, Chuo-ku, Sagamihara, Kanagawa, Japan}
\affil{5 INAF Osservatorio di Radioastronomia, via Gobetti 101, 40129 Bologna, Italy }
\affil{6 INAF - Istituto di Astrofisica e Planetologia Spaziali di Roma (IAPS), Via del Fosso del Cavaliere 100, 00133 Roma, Italy}
\affil{7 SRON Netherlands Institute for Space Research, Sorbonnelaan 2, 3584 CA Utrecht, the Netherlands}

\begin{abstract}
Winds outflowing from Active Galactic Nuclei (AGNs) may carry significant amount of mass and energy out to their host galaxies. 
In this paper we report the  detection of a sub-relativistic outflow observed in the Narrow Line Seyfert~1 Galaxy IRAS17020+4544 as a series of absorption lines corresponding to at least 5 absorption components with  an unprecedented wide range of associated column densities and ionization levels and velocities in the range of  23,000-33,000~km/s, detected at X-ray high spectral resolution ($E / \Delta E$$\sim$1000) with the ESA's observatory XMM-Newton. The charge states of the material constituting the wind clearly indicate a range of low to moderate ionization states in the outflowing gas and column densities significantly lower  than observed in highly ionized ultra fast outflows. 
We estimate that at least one of the outflow components may carry sufficient energy to substantially suppress star formation, and heat the gas in the host galaxy. IRAS17020+4544 provides therefore  an interesting example of feedback by a moderately luminous AGN hosted in a spiral galaxy, a case barely envisaged in most evolution models, which often predict that feedback processes take place in massive elliptical galaxies hosting luminous quasars in a post merger phase.
  \end{abstract}
  

\section{Introduction}
The effect of  AGN winds on the larger scale environment,  which is commonly referred to as ``AGN feedback",  has far-reaching effects on the evolution of the host galaxy properties (Di Matteo et al. 2005, Hopkins et al. 2010). For this reason, quasar feedback has been extensively postulated as a fundamental mechanism for regulating the relation of supermassive black holes with their host galaxies. 
The recent reports of sub-relativistic X-ray winds in two very bright quasars (Tombesi et al. 2015, Nardini et al. 2015) corroborates the role of highly ionized accretion disc winds as a source of feedback. 
Indeed, the X-ray spectra of some AGNs  carry the signature of gas outflowing at sub-relativistic speed ($v \ge 0.1$~c), so highly ionized that the only dominant ions are He-like and  Hydrogen-like ionic species. These systems are known as ``Ultra-Fast Outflows" (UFOs) and they are observed in the Fe K band at E$\ge$~7~keV.
Several papers reported on UFOs hosted in individual AGNs (Pounds et al. 2003,2011,2014, Chartas et al. 2009, Lanzuisi et al. 2012), and statistical studies show that UFOs are detected in 30-40\% of nearby AGN (Tombesi et al. 2010, 2012, Gofford et al. 2013).  The approximate ranges of mass outflow rate (0.01-1M$_{\odot}$~yr$^{-1}$) and kinetic energy (10$^{42-45}$ erg~s$^{-1}$) may be in good agreement with theoretical predictions for black hole winds (King 2010). 
   However, the  very existence of such winds has been debated within the community (Laha et al. 2014, Kaspi et al. 2006, Gallo et al. 2013, Zoghbi et al. 2015) because the intrinsic limitations in the CCD data resolution ($E / \Delta E$$\sim$30-40) and the often uncertain identification of the atomic transition do not allow an unambiguous -model independent- confirmation that these features represent a bona-fide UFO. 
 
IRAS17020+4544 is a Narrow Line Seyfert~1 spiral Galaxy at redshift {\it z}=0.0604 with black hole mass of M$_{BH}$$\sim$5.9$\times$10$^{6}$M$_{\odot}$ (Wang et al. 2001), X-ray luminosity of $\sim$1.5$\times$10$^{44}$~ergs~s$^{-1}$ and flux of 1.6$\times$10$^{-11}$~ergs~cm$^{-2}$~s$^{-1}$ in 0.3-10 keV. The bolometric luminosity of $\sim$5.2$\times$10$^{44}$~ergs~s$^{-1}$ estimated assuming a conservative bolometric correction {\it k}=10 (Marconi et al. 2004)  implies that the source is accreting at a high accretion rate  (L$_{\rm BOL}$~/~L$_{\rm EDD}$ $\sim$0.7). 

\section{Data reduction and spectral analysis}
XMM-Newton observed IRAS17020+4544  for 160~ks in 2014 on January 23rd and 25th  (OBSIDs 0721220101 0721220301), and twice  in 2004 for 40~ks  (0206860101 0206860201).
The X-ray data  include the spectra from the Reflection Grating Spectrometer (RGS) (den Herder et al. 2001), and the CCD spectra from the European Photon Imaging Camera (EPIC pn) (Struder et al. 2001).
No background flares due to high-energy particles are present in the observations. 
RGS data were processed by the standard System Analysis Software (SAS 13.5.0) tool {\it rgsproc}. Due to the minor flux variability and the lack of spectral changes within the two XMM OBSIDs of each year ($\Delta\Gamma$=3\%, $\Delta$Flux=9\%), RGS spectra were combined in one single spectrum  by the SAS tool {\it rgscombine} in order to maximize the S/N. EPIC pn event files were obtained  by applying standard filtering criteria through the SAS task {\it epproc}. 
Source and background spectra were extracted from circular regions of 40$^{\prime\prime}$. Background regions were selected in source-free portions of the CCD. Response matrices were provided by the SAS tools {\it arfgen} and {\it rmfgen}.
The software used for the spectral analysis is XSPEC v.12.8.2h. 

\subsection{The continuum and the warm absorbers}
The RGS spectral analysis was carried out on the unbinned spectrum and the Cash statistics (Cash 1979) was applied in the spectral fitting process.  Fit parameters are quoted with 1$\sigma$ error bars throughout the paper. The X-ray continuum in 7-35~\AA~ was parametrized by a power law modified by Galactic absorption (Kalberla et al. 2005) of column density N$_{\rm H}$=2$\times$10$^{20}$~cm$^{-2}$.  The residuals of the spectrum against the best-fit continuum show a forest of narrow absorption features in the soft X-ray band imprinted by ionized gas along the line-of-sight to the AGN. We first modeled the warm absorber  that was already known in this source (Komossa et al. 1998). To this purpose we employed the self-consistent photoionization code PHASE  (Krongold et al. 2003). Each PHASE component describes one layer of ionized gas characterized by a particular set of parameters: ionization state U\footnote{$U = \frac{Q}{(4 \pi R^2 c n_e)}$ where Q is the luminosity of ionizing photons, n$_e$ and R are the electron density and the distance of the wind from the X-ray source, respectively}, column density, covering factor, turbulent velocity and outflow velocity. To compute the absorption spectrum we assumed the ionization balance produced by the incoming ionizing radiation emitted by the source, therefore we constructed the Spectral Energy Distribution (SED) for IRAS17020+4544 using data from the Nasa Extragalactic Database and from XMM-Newton. For fitting the warm absorber  four such absorbing components were required by the data (Longinotti et al. in prep). 
 The velocity component along the line of sight of the 4 absorbers is found of the order of 10$^{2-3}$~km/s.
  The photon index of the power law is $\Gamma$=2.8$\pm$0.04, as in typical Narrow Line Seyfert 1 sources. 
The model including the power law continuum and the 4 ionized absorbers yields a fit statistics of C$_{stat}$=3258 for 2752 degrees of freedom in the 7-35~\AA~band (Figure 1). 
\begin{figure*}
\centering
\includegraphics[width=1.8\columnwidth, height=1.18\columnwidth]{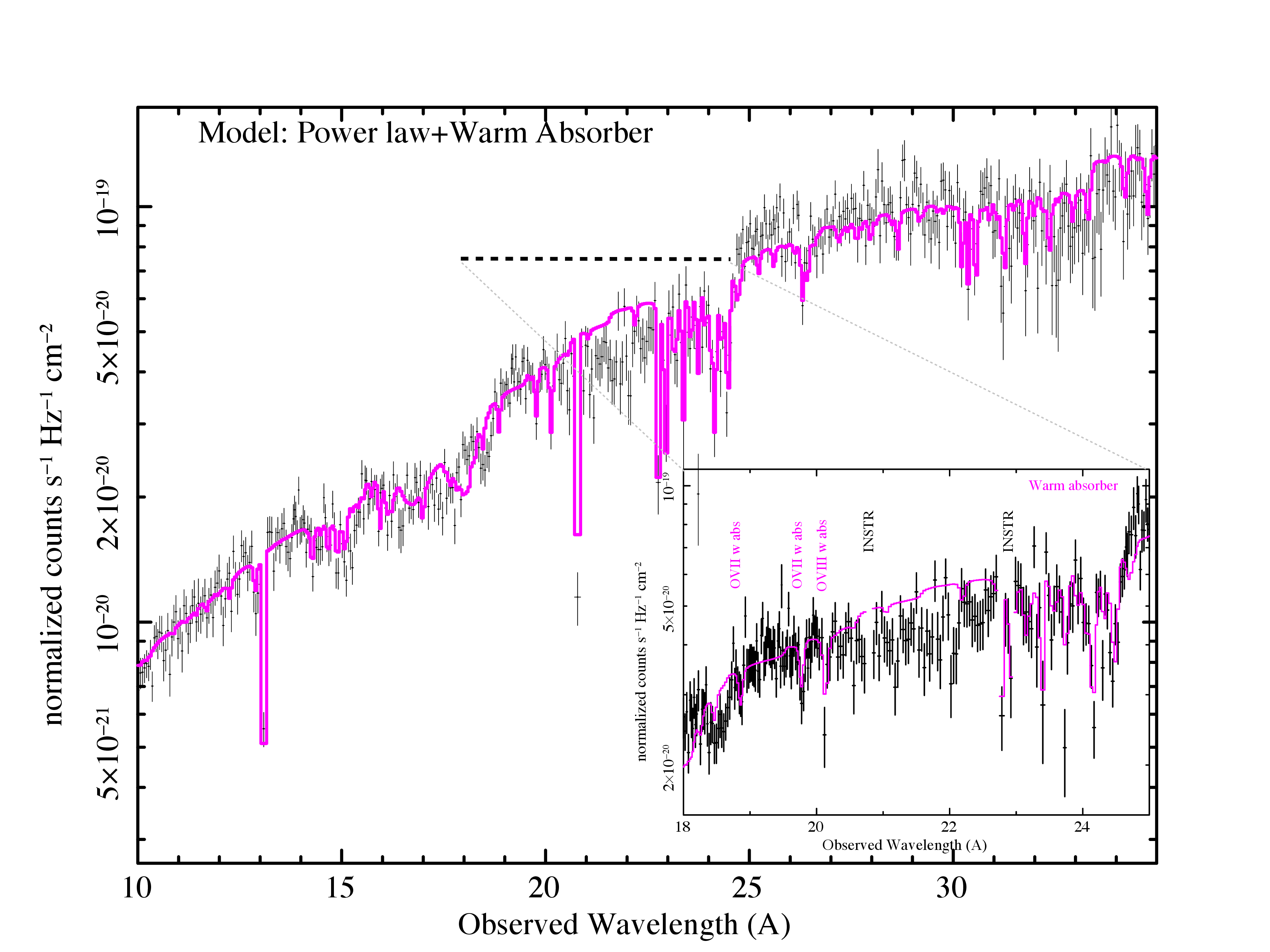}
\caption{RGS spectra fitted with the model including a power law continuum absorbed by 4 ``slow" warm absorber components.  
The inset shows the spectral portion mostly affected by the ultra fast outflow. The spectrum was binned by 5 instrumental channels for display purposes.}
\end{figure*}

Once the warm absorber was well parametrized we noticed the presence of additional absorption features, particularly pronounced in the 18-23~\AA~band (see inset of Figure 1). 
Initially, we  tested for the presence of individual absorption lines by adding narrow Gaussians lines with negative intensity to the continuum+warm absorbers model.  The line widths are fixed to 0.1~eV and their position is free.  
Table 1 reports the list of the absorption lines detected above a 2$\sigma$ threshold, along with their positions and the identification of the atomic transition. 
The improvement in C-statistics is measured for the addition of 2 d.o.f. and in the spectral band 7-35~\AA~. 
The results reported in Table 1 show that additional absorption is required by the data and, supported by this evidence, we proceed to model it in a self-consistent manner.

\begin{figure}[t]
\centering 
\includegraphics[width=0.8\columnwidth, height=1.1\columnwidth,angle=-90]{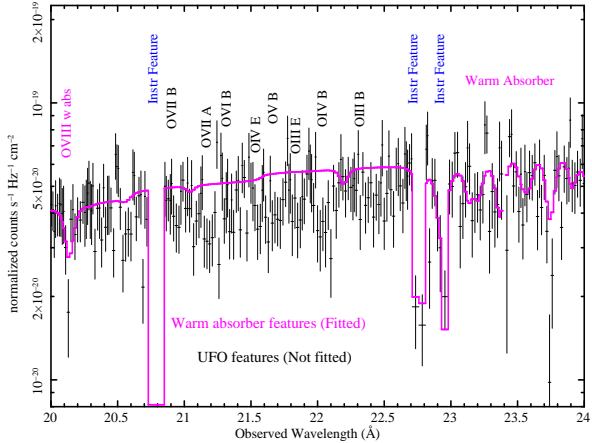}
\caption{Features from the warm absorber (included in the model) and from the UFO (not included) do not overlap with each other. Instrumental features (in blue)  do not hamper the detection of absorption  intrinsic to the source. Labels are drawn from Table 2 physical model.}
\end{figure}

\subsection{Modeling the Ultra Fast Outflow}
The analysis was then restricted to the spectral region where residuals are most prominent (18-23\AA) and that is the least contaminated by the warm absorber (see Figure 1), which imprints only two unblended absorption lines from OVII and OVIII in this spectral range. The statistics for the model with a local power law continuum and the 4 warm absorbers fitted in this band is  C$_{stat}$/d.o.f.=512/478.  
 We start by adding to this model a fast outflow component  modeled by PHASE. Four additional such components were included in the model. After validating the statistical improvement of each of them, five UFO components are required by the fit (see  Table~2). 
In contrast to adding Gaussian lines at random positions that are not descriptive of the global ionization state of the gas, the use of  PHASE  guarantees that the absorption is modeled self consistently.  The lines are narrow and unresolved by RGS, so the turbulent velocity was fixed to 50 km/s. All UFO components have outflow velocities consistent with the lines being displaced by {\it v}$\sim$0.1{\it c}. 
The final fit statistics for the best fit model that includes the power law continuum, the 4 warm absorber components plus 5 UFO components is C$_{stat}$/d.o.f.=409/463. 
Figure 2 shows  that our analysis is fully capable to separate the two sets of lines from the slow and fast outflowing gas without ambiguity. 
When the best fit model is extended to the full RGS band down to 35~\AA, the detection of the UFOs is confirmed, therefore we are confident that these features cannot be mistaken with the absorption lines from the warm absorber at lower wavelengths. We also tested the robustness of the UFO detection by estimating its significance using a local continuum in the 18-23 Angstrom band that does not include the warm absorber. All five components are detected with the same significance. Figure 3 shows the RGS spectrum fitted by the best fit model. 

Although an alternative scenario is unlikely since the pattern of the line series observed in our data are very well fitted by our self-consistent model, we have considered the presence of local absorption ({\it z}=0) as possible alternative to the ultra fast wind. Fitting a PHASE component with velocity outflow fixed within a range consistent with the absorption being local  did not provide a satisfactory fit to the spectrum as the positions of the absorption lines if interpreted at {\it z}=0, are not consistent with known ionic transitions.
In particular, most of the local low-ionization absorption (Gatuzz et al. 2013) at  {\it z}=0 lies in the range 22-24~\AA, i.e. above the region where the bulk of the UFOs lines are detected. 
Likewise, the absorption lines cannot be attributed to intervening material between us and the galaxy, as the measured outflow velocity is significantly larger than the receding velocity of IRAS17020+4544 (18,000~km~s$^{-1}$).
\begin{table}[t]
\centering
\caption{List of absorption lines individually detected above a 2$\sigma$  threshold. Identifications were made by matching this phenomenological fit  to the physical model in Table 2.}
\scriptsize
\begin{tabular}{c | c | c | c | c}
\\
\hline\hline
Obs $\lambda$  &  Intensity &  $\Delta$Cstat   &  Sign. &   Line \\ 
  (\AA)   &  (10$^{-5}$ ph~cm$^{-2}$~s$^{-1}$)     &    -            &               $\sigma$  &   ID   \\
\hline\hline 
   21.18$^{+0.02}_{-0.02}$               &  -2.7$^{+0.60}_{-0.60}$           &  34     &  4.5     &    OVII \\ 
   22.03$^{+0.04}_{-0.02}$   & -3.01$^{+0.89}_{-0.73}$  &  32     &  3.4   & OIV \\ 
    18.60$^{+0.03}_{-0.03}$                & -1.54$^{+0.55}_{-0.55}$  & 19     & 2.8  & OVIII  \\ 
   21.67$^{+0.03}_{-0.03}$                & -2.17$^{+0.80}_{-0.80}$              &     21  & 2.7  &   OV \\ 
  21.83$^{+0.03}_{-0.01}$   &    -2.28$^{+0.92}_{-0.92}$            &  18     &  2.5  &  OIII   \\ 
   21.55$^{+0.03}_{-0.02}$   & -1.76$^{+0.85}_{-0.91}$  &   8      & 2.0 & OIV   \\    
   20.93$^{+0.02}_{-0.03}$   & -1.61$^{+0.83}_{-0.82}$  &  10    & 2.0   & OVI \\ 
  \hline 
\end{tabular}
\end{table} 
\subsubsection {XMM-Newton data of 2004} 
We tested the presence of the fast outflow in XMM-Newton archival data of IRAS 17020+4544 obtained in 2004. We kept the baseline model with the same continuum and the warm absorber as the overall spectral shape and flux level in 2004 are very similar to those inferred in  2014. 
This yielded a fit statistics of C$_{stat}$/d.o.f.=514/485  in 18--23~\AA~. When the ultra fast outflow was added to the model, only component A (i.e. the most significant in the 2014 data) is detected in this data set, with an improvement of $\Delta$C$_{stat}$=21. The parameters of this outflow component are consistent with the ones reported for 2014.  The other four components are not detected but we cannot statistically exclude their presence in these data. This test shows that at least one component (A)  of the outflow seems persistent over a 10~yr time scale.
 
\begin{table*}
\centering 
\caption{ Parameters of the 5 UFO components detected in the RGS spectrum.  The statistical improvement (5th column) refers to the addition of each PHASE component to the model comprising the continuum, the warm absorbers and the previous UFO components.
The significance is estimated through Monte Carlo methods.}
\begin{tabular}{c | c | c | c | c| c}

\hline\hline
UFO Component & log U        & Log N$_{\rm H}$     &  $\rm v_{out}$  & Statistics  & Significance \\
         
 index       & (erg cm s$^{-1}$) & (cm$^{-2}$)  & (km/s) &  $\Delta$C$_{stat}$  & \\
\hline 
 Comp A)    &   -0.39$^{+0.30}_{-0.15}$  & 21.47$^{+0.18}_{-0.21}$  &   23640$^{+150}_{-60}$    &  45 &  9.0$\sigma$ \\ 
   
 Comp  B)   &  -1.99$^{+0.33}_{-0.26}$   &  20.42$^{+0.21}_{-0.58}$  &  27200$^{+240}_{-240}$   &  26 & 5.3$\sigma$ \\ 
 
 Comp  C)   &   2.58$^{+0.17}_{-0.85}$   & 23.99$^{+0}_{-1.86}$   &  27200$^{+300}_{-270}$        &   10 & 3.6$\sigma$    \\ 
 
 Comp  D)   &   0.33$^{+1.79}_{-0.40}$   & 21.42$^{+0.84}_{-1.28}$   & 25300$^{+210}_{-180}$    &   12 & 2.6$\sigma$ \\
   
 Comp  E)   &   -2.92$^{+0.51}_{-0.14}$  & 19.67$^{+0.34}_{-0.36}$   &   33900$^{+360}_{-270}$  &    10 &  2.0$\sigma$  \\ 
\hline
\end{tabular}
\end{table*}

\subsection{Monte Carlo Methods}
To provide a rigorous estimate of the statistical significance of the  outflow detected in 2014 Monte Carlo simulations were performed. 
Following the approach that was taken for modeling the UFO in the RGS spectra, the restricted band 18-23~\AA~was considered.  
As a baseline model for the synthetic data sets we assumed the (continuum+ 4 warm absorbers) model for testing the significance of Component A. 
For the other UFO components we always included all the previous UFO components in the model as done in the real spectral fitting, i.e. we included  (continuum+ 4 warm absorbers+UFO A) for testing the significance of UFO B, (continuum+ 4 warm absorbers+UFO A+UFO B) for testing the significance of UFO C, etc. 
For each UFO we produced 1000 simulated spectra with the XSPEC package. Each realization was folded through the same response matrix employed in the spectral fitting, and random noise was added, therefore each spectrum corresponds to a background-subtracted RGS data set with same photon statistics as the real data set. In each simulated spectrum the improvement in C-statistics for adding a UFO component to the baseline model was measured and recorded. We checked that the results were distributed according to a Gaussian distribution and this conclusion led to the significance quoted in Table~2. The significance of Component C was estimated by tying the wind velocity between Components B and C to be the same within the errors. Given the two very similar outflow velocities, the information on the velocity of Component B is used as a prior for assessing the significance of Component C. 
\subsection{EPIC pn spectral analysis}
 We searched for a potential high-ionization counterpart of the RGS UFO in the 2014 EPIC-pn spectrum. The data fitted in the 2-10 keV band with a  steep power law continuum  ($\Gamma$=2.25$\pm$0.08) show a prominent Fe K$\alpha$ line in emission that was fitted phenomenologically with a Gaussian profile. The resulting peak energy is consistent with emission from ionized material (E=6.95$\pm$0.10keV), the line width is $\sigma$=0.26$^{+0.21}_{-0.10}$~keV and the Equivalent Width is 150$^{+85}_{-55}$~eV. We then added a negative narrow Gaussian line at the position of the Fe~XXVI atomic transition (6.97 keV rest frame) blue-shifted by the corresponding velocity measured in the RGS UFO (i.e. E$\sim$7.2 keV). No absorption line is detected at this position. The 90\% upper limit on the line EW is 14 eV is significantly lower than the average intensity of Fe~K  UFOs  (Tombesi et al. 2010). We then explored the potential presence of absorption lines at higher energies by adding to the power law continuum two narrow Gaussians with free position. 
Two features are marginally detected at E=8.75$\pm$0.13~keV and E=9.39$\pm$0.10 keV, with a respective intensity of 49$\pm$43~eV  and EW=64$\pm$50 eV. Although the signal-to-noise in this region of the EPIC-pn spectrum is low due to the sharp decrease of the XMM-Newton telescopes' effective area above 7 keV and to the steep intrinsic spectrum of IRAS17020+4544, 
 the positions of the two absorption lines are consistent with being Fe~XXV and Fe~XXVI blue-shifted by a velocity of {\it v}$\sim$0.34{\it c}.

\begin{figure}[t]
\centering
   \includegraphics[width=0.8\columnwidth, height=1.1\columnwidth,angle=-90]{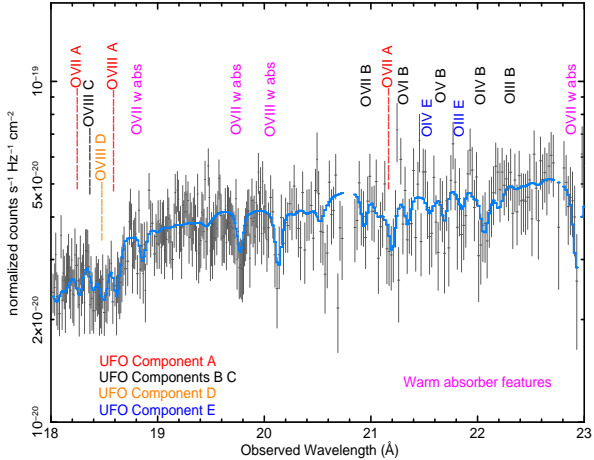}
\caption{ Portion of the RGS spectrum mostly affected by the fast wind absorption and modeled by the best fit reported in Table 2. }
\end{figure} 

\section{Results}
\subsection{A stratified multi-component ultra fast outflow}
The simultaneous presence of three UFO components required with such a high significance (9.0, 5.3 and 3.6$\sigma$) is, to our knowledge, unprecedented. More importantly,  the newly discovered wind presents a much more complex nature than envisaged so far for UFOs. Components B and C are separated by four decades in ionizations and more than three decades in column density, yet they are outflowing at the same speed.

Most theoretical models of black hole winds predict that winds at sub-relativistic speed  are bound to reach a ionization level so high as to produce only absorption in the Fe K band, e.g. the shock outflow model (King 2010).
However, the unquestionable presence of a sub-relativistic wind at low-ionization in IRAS17020+4544 breaks the linear correlation between ionization and velocity of the wind that is predicted by this model and that inspired  the unification scenario for UFO and warm absorbers  postulated by Tombesi et al. (2013). We note that  this scenario has already been questioned (Gupta et al. 2013, 2015).

The  newly discovered multi-component outflow could be interpreted as gas condensed in clumps with distinct properties ({\it v}, U, N$_{\rm H}$) embedded in a high-velocity, high-ionization wind as the one described in the shock outflow model. However,  it would be difficult to explain how these clumps may survive for t$\ge$10~yr, as inferred by the comparison of the XMM-Newton spectra in 2014 and 2004. More likely, the observed  absorption lines are revealing the cooling region envisaged  in models of black hole outflows right after the first shock between the wind and the surrounding material (King et al. 2010). In this model, the accretion disc wind launched with velocity v$_{out}$ suffers an isothermal shock with the circumnuclear gas, which produces the effect of slowing the wind to a velocity {\it v} $\sim$$\frac{v_{out}}{4}$. The properties of the two tentative highly ionized absorption lines reported in Section 2.4 are consistent with the predictions of this model. 
Although the structure of the outflowing system in  IRAS17020+4544 requires a more elaborate explanation, the observed velocities of the wind seem in excellent agreement with this scenario:  {\it v$_{out}$}$\sim$0.34{\it c}=102,000~km~s$^{-1}$ (EPIC pn measurement) and v$\sim$0.08-0.11{\it c}=24,000-33,000~km~s$^{-1}$ (RGS measurement).  Deeper X-ray observations will certainly help to shed light on the presence  and properties of the highly ionized features.

  An additional point of interest of the outflow in IRAS17020+4544 comes from observations in the radio band. Recent data obtained by our group (Giroletti et al. in prep) along with a re-analysis of archival data (Gu et al. 2010) confirm the presence of outflowing material on a scale of about 10~pc with a speed lower than $\sim$0.2{\it c}, potentially in good agreement with the X-ray velocities. The presence of the radio jet/outflow suggests that the wind may be magnetically driven or produced by the interaction of the radio jet with the surrounding material. On the other hand,  the high accretion rate of the source (L$_{\rm BOL}$~/~L$_{\rm EDD}$ $\sim$0.7)  is in agreement with models of radiatively driven outflows (King et al. 2010).

\subsubsection{Comparison with other results}
Detection of ultra fast outflows in high-resolution spectra of local AGN  with moderate ionization was already reported.
  Pounds (2014) reports a secure detection of only one component  in the RGS data of  PG1211+143 since the second outflow may well be associated to local absorption due to the coincidence of its velocity with the systemic velocity of the AGN. 
The claim of a dual relativistic outflow in Ark~564 (Gupta et al. 2013) is based on measurement of individual lines with velocities inconsistent with each other, therefore it is difficult to associate them to the same outflowing system.  Finally, only one absorber  was detected above the threshold of 99.99\% in the Chandra grating spectra of Mrk 590  (Gupta et al. 2015). 
While all these results have certainly contributed to foster the search for UFOs, none of them unveiled an ultra fast outflow constituted  by several ionization and velocity components like the one discovered in IRAS17020+4544. 
Our result represents therefore the first unambiguous discovery of a fast (v$\sim$0.1~{\it c})  multi-component AGN outflow in X-ray high-resolution spectroscopic data, ultimately settling the open controversy on their existence.

\subsection{Energetics of the ultra fast outflow and consequences for feedback}
The energetics of the outflow were estimated under the common assumption that the outflow velocity is larger or equal to the escape velocity at the launch radius $r=\frac{2GM_{BH}}{v^2_{out}}$. Taking into account the lowest and highest velocities measured in the wind (corresponding to components A and E, respectively) the launching radius is limited to be lower than  (1.4--2.8)$\times$10$^{14}$~cm.
Following Tombesi et al. (2015) and Crenshaw et al. (2012) the expression $\dot{M}_{{out}}$=4$\pi$$\mu$rN$_{\rm H}$$v_{out}m_p$C$_f$  can be used for estimating the mass outflow rate in spherical geometry with {\it m$_p$} and {\it $\mu$} corresponding to the mass of the proton and to the mean atomic mass per particle ({\it $\mu$}=1.4), C$_f$ as the covering fraction of the  outflow, and N$_{\rm H}$ and v$_{out}$ as the column density and the velocity of the flow measured in our data. 
 By expressing these two quantities respectively in units of 10$^{21}$~cm$^{-2}$ and 10,000 km~s$^{-1}$, this formula can be conveniently parametrized as $\dot{M}_{{out}}$=7.3$\times$10$^{-4}$~$\frac{M_{\odot}}{yr}$~$\frac{N_H21}{v_{out(10,000)}}$~C$_f$ gr~s$^{-1}$, which yields a relevant mass outflow rate only for Component C, i.e. $\dot{M}_{{out}}$(C) $\sim$0.26~C$_f$~$M_{\odot}$yr$^{-1}$.
 The energy outflow rate can be estimated starting by $\dot{E}$=$\frac{1}{2}$$\dot{M}_{{out}}$v$^{2}_{out}$ where the previous expression for $\dot{M}_{{out}}$ can be substituted in order to express the energy outflow rate in terms of the column density and the velocity of the wind: $\dot{E}$=2.31$\times$10$^{40}$~N$_{H21}$ v$_{out}$(10,000)~C$_f$~erg~s$^{-1}$.
Considering the bolometric luminosity, we can express $\frac{\dot{E}}{L_{BOL}}$= $\frac{\dot{E}}{{\it k}L_{2-10}}$, which for Component C yields  $\frac{\dot{E}(C)}{L_{BOL}}$=11\%C$_f$~erg~s$^{-1}$.  

The study of the properties of this outflow may hold important clues for elucidating the contribution of accretion disc winds to AGN feedback.
Even if the outflow covers a small part of the solid angle, the energy rate is sufficient to power AGN feedback according to the  required minimum value (Hopkins et al. 2010)  of $\sim$0.5\%. Most notably, the AGN luminosity in IRAS17020+4544 is $\sim$100 times lower than in the sources where the most compelling evidence for black hole feedback produced by X-ray highly ionized fast winds was found (Tombesi et al. 2015, Nardini et al. 2015). Deep optical images (Ohta et al. 2007) show that this AGN is hosted by a spiral barred galaxy with no signs of interactions. 
 Hence, IRAS17020+4544 is the first case of an undisturbed field spiral galaxy where feedback from accretion disk winds can regulate star formation and drive the fundamental relation between the stellar velocity dispersion in the galaxy bulge  and the mass of the central supermassive black holes (Ferrarese et al. 2000).
These findings will certainly offer new insights to reveal the physical mechanisms at play during the phase when galaxy evolution is regulated by black hole feedback.

\acknowledgments
This paper is based on observations obtained with XMM-Newton, an ESA science mission with instruments and contributions directly funded by ESA Member States and NASA. We thank  the anonymous referee. We acknowledge support from the ESAC Faculty, INTEGRAL ASI/INAF n. 2013-025.R.O, CONACYT n.233263. 
Y.K aknowledges support from grant DGAPA PAIIPIT IN104215 and CONACYT grant 168519. This paper is dedicated to Prof. Maurizio~Longinotti, Haematologist.

\end{document}